# A Governance Model for IoT Data in Global Manufacturing

Vignesh Alagappan
Rheem Manufacturing, Roswell, Georgia, USA,
Vignesh.alagappan@rheem.com

*Abstract*—Industrial IoT platforms in global manufacturing environments generate continuous operational data across production assets, utilities, and connected products. While data ingestion and storage capabilities have matured significantly, enterprises continue to face systemic challenges in governing IoT data at scale. These challenges are not rooted in tooling limitations but in the absence of a governance model that aligns with the realities of distributed operational ownership, heterogeneous source systems, and continuous change at the edge. This paper presents a federated governance model that emphasizes contract-driven interoperability, policy-as-code enforcement, and asset-centric accountability across global manufacturing organizations. The model addresses governance enforcement at architectural boundaries, enabling semantic consistency, quality assurance, and regulatory compliance without requiring centralized control of operational technology systems. This work contributes a systems architecture and design framework grounded in analysis of manufacturing IoT requirements and constraints; empirical validation remains future work

*Keywords*—Internet of Things, data governance, manufacturing systems, industrial IoT, federated governance, data contracts, policy-as-code, system architecture

## I. INTRODUCTION

Industrial IoT platforms in global manufacturing environments generate continuous operational data across production assets, utilities, and connected products. While data ingestion and storage capabilities have matured significantly, enterprises continue to face systemic challenges in governing IoT data at scale. These challenges are not rooted in tooling limitations, but in the absence of a governance model that aligns with the realities of distributed operational ownership, heterogeneous source systems, and continuous change at the edge.

IoT data differs fundamentally from traditional enterprise data in four critical dimensions. First, it is event-driven rather than transaction-based, with temporal ordering and correlation requirements that static relational models cannot express. Second, it is asset-bound, meaning every signal carries implicit context about physical equipment, location, and operational state. Third, it is produced by systems not centrally governed by enterprise IT, including programmable logic controllers (PLCs), supervisory control and data acquisition (SCADA) systems, and vendor-specific historians that operate under operational technology (OT) ownership. Fourth, it exhibits high temporal variability, with sample rates ranging from milliseconds to hours depending on process criticality and equipment characteristics

As a result, governance mechanisms that rely on centralized ownership, static schemas, or post-ingestion controls fail to scale. Data quality issues propagate from edge to analytics without clear accountability. Schema changes at the source break downstream pipelines without warning. Regulatory requirements vary by geography and product line, fragmenting architectures into compliance silos. Cross-domain data reuse stalls due to semantic ambiguity and unclear ownership.

This paper defines a governance model for IoT data that emphasizes contract-driven interoperability, policy-as-code enforcement, and federated accountability across global manufacturing organizations. The model separates policy definition from implementation and enforcement, distributes authority while centralizing compliance verification, and operates at architectural boundaries rather than within operational systems

## II. RELATED WORK

IoT data governance has been approached from multiple perspectives, including centralized cloud architectures, federated data mesh patterns, and industry-specific reference models. This section positions our federated governance model against existing approaches, highlighting key differences in governance philosophy, architectural assumptions, and operational implementation

### A. Data Mesh Architectures

Data mesh represents a paradigm shift from centralized data platforms to domain-oriented decentralization [1]. Dehghani's foundational work establishes four principles: domain ownership, data as a product, self-serve data infrastructure, and federated computational governance. Data mesh architectures treat data products as first-class entities with explicit contracts, quality guarantees, and discoverable interfaces.

Our model adopts data mesh principles while addressing manufacturing-specific constraints that distinguish IoT governance from enterprise analytics. First, data mesh assumes domains can independently manage source systems, whereas manufacturing IoT domains consume data from operationally controlled assets (PLCs, SCADA) that cannot be modified to accommodate governance requirements. Second, data mesh emphasizes streaming and batch analytics workloads, while manufacturing requires real-time governance decisions at ingestion boundaries without blocking operational data flows. Third, data mesh governance focuses on dataset-level policies, whereas IoT requires asset-centric governance that references physical equipment attributes and lifecycle states.

Our contribution extends data mesh with boundary-enforced governance at OT-to-IT transitions. These asset identity models enable fine-grained policy application and an explicit separation of plant-level operational execution from domain-level semantic ownership. These extensions address the operational reality that manufacturing plants cannot implement domain governance requirements directly on source systems

### B. Industrial Internet Consortium Reference Architecture

The Industrial Internet Consortium (IIC) Industrial Internet Reference Architecture (IIRA) provides a comprehensive framework for industrial IoT systems spanning edge, platform, enterprise, and control tiers [9], [10]. IIRA defines four architectural viewpoints: business, usage, functional, and implementation. The functional view includes data management as a core component, covering collection, ingestion, transformation, and analytics.

IIRA's data management model emphasizes data lifecycle (acquisition, transport, storage, analysis, visualization) and quality attributes (accuracy, completeness, consistency, timeliness, validity). However, IIRA does not prescribe specific governance mechanisms for enforcing these attributes at scale across federated domains. The reference architecture describes what should be governed, but not how governance is operationalized across organizational boundaries.

Our model complements IIRA by providing concrete governance mechanisms—data contracts, policy-as-code, boundary enforcement—that implement IIRA's data quality principles. Where IIRA defines functional requirements, our model defines architectural patterns and enforcement points. Specifically, we operationalize IIRA's data quality attributes through contract-based validation at ingestion boundaries, continuous quality monitoring with automated remediation, and federated accountability structures that assign ownership without requiring centralized control.

The key distinction is governance operationalization: IIRA provides a descriptive framework, while our model provides prescriptive implementation patterns for manufacturing environments.

### C. Cloud Provider IoT Governance Models

Major cloud providers offer IoT platforms with integrated governance capabilities: AWS IoT Core with AWS Glue Data Catalog, Azure IoT Hub with Azure Purview, and Google Cloud IoT Core with Dataplex. These platforms provide device management, data ingestion, transformation pipelines, and metadata cataloging as managed services.

Cloud provider models assume governance is implemented through centralized platform services controlled by enterprise IT. Device registration, data routing, schema enforcement, and access policies are configured through provider-specific APIs and consoles. Governance is enforced within the cloud provider's control plane, with limited visibility into edge and plant systems

This approach creates several challenges for global manufacturing. First, vendor lock-in constrains migration flexibility as governance configuration is tightly coupled to proprietary APIs. Second, centralized enforcement requires reliable network connectivity between plants and the cloud, conflicting with operational requirements for network segmentation and local autonomy. Third, governance policies are expressed in provider-specific languages (IAM policies, Azure RBAC, Google IAM) that do not translate across platforms, preventing multi-cloud strategies.

Our federated model addresses these limitations through provider-agnostic governance mechanisms. Data contracts are defined in open formats (JSON Schema, Avro, OWL) independent of cloud platforms. Policy-as-code uses standard languages (Open Policy Agent, Cedar) that evaluate consistently across environments. Boundary enforcement occurs at architectural transitions (OT-to-IT, raw-to-canonical) rather than within cloud provider control planes, enabling hybrid and multi-cloud deployments.

The critical distinction is the location of governance: cloud providers enforce governance within their platforms, while our model enforces governance at domain boundaries regardless of deployment topology. This enables plant-level autonomy with centralized compliance verification, addressing manufacturing's operational constraints.

### D. Data Spaces and Industry 4.0 Initiatives

European data space initiatives, notably the International Data Spaces Association (IDSA) Reference Architecture Model and Gaia-X framework, define governance for cross-organizational data sharing [28], [31], [32]. Data spaces establish trust through certified connectors, usage control policies, and decentralized identity management. These initiatives address data sovereignty, competitive sensitivity, and regulatory compliance in multi-party manufacturing ecosystems.

Data spaces focus on inter-organizational governance (sharing data between companies) while our model addresses intra-organizational governance (governing data across plants and domains within an enterprise). The governance challenges differ: data spaces prioritize contractual agreements and bilateral trust, while enterprise IoT governance requires operational scalability and real-time enforcement across thousands of assets.

However, data space concepts inform our model's boundary enforcement approach. IDSA connectors enforce usage policies at data egress points, similar to our enforcement at architectural boundaries. Our contribution adapts these patterns to manufacturing IoT scale, where millions of data points per second require automated policy evaluation without bilateral negotiation overhead

### E. Positioning and Contributions

Existing approaches address subsets of IoT governance challenges but do not comprehensively solve federated governance in global manufacturing:

- Data mesh provides domain ownership principles but assumes domain control over source systems, which

manufacturing cannot achieve due to operational constraints.

- IIRA defines what should be governed but not how governance is operationalized across federated domains with heterogeneous source systems.
- Cloud providers offer managed governance within their platforms but create vendor lock-in and assume centralized cloud control incompatible with manufacturing's network segmentation and local autonomy requirements.
- Data spaces address inter-organizational trust but do not scale to high-volume, real-time intra-organizational IoT governance.

Our model contributes:

- Boundary-enforced governance that respects operational autonomy while ensuring compliance, addressing the reality that manufacturing cannot modify source systems to accommodate governance requirements.
- Asset-centric policy application enabling fine-grained governance without schema fragmentation, leveraging physical asset hierarchies and lifecycle states.
- Federated accountability structures that separate policy definition (enterprise), interpretation (domain), and execution (plant) without centralized bottlenecks.
- Provider-agnostic mechanisms using open standards (Avro, OWL, OPA) that function consistently across hybrid and multi-cloud deployments.

These contributions enable governance at manufacturing IoT scale while preserving the operational autonomy required for production continuity.

### III. PROBLEM DEFINITION AND SCOPE

*A. Core Governance Problem*

The core governance problem in IoT platforms is the inability to consistently establish trust, reuse, and accountability across data produced by independent operational domains [22], [25]. Manufacturing plants operate with varying automation maturity, vendor ecosystems, and local constraints. Changes at the source—such as PLC logic updates, firmware revisions, or asset replacements—frequently occur without coordination with downstream consumers.

This creates three failure modes. First, semantic drift occurs when signal definitions change at the source without updating metadata or notifying consumers. A temperature sensor recalibrated from Fahrenheit to Celsius breaks analytics pipelines that assume unit consistency. Second, quality degradation happens when equipment malfunctions or network issues introduce data loss, duplication, or timestamp corruption that downstream systems cannot detect or remediate. Third, compliance violations emerge when data handling does not align with regulatory requirements for retention, access control, or cross-border transfer.

Traditional data governance frameworks assume centralized ownership, stable schemas, and batch-oriented ingestion. These assumptions do not hold in manufacturing IoT environments where operational domains control data production, schemas evolve continuously, and streaming architectures require real-time governance decisions.

*B. Scope Definition*

The scope of this paper includes:

- Governance of operational IoT data from edge to cloud
- Semantic consistency, quality assurance, access control, and lifecycle governance
- Organizational governance across enterprise, domain, and plant layers
- Contract-based interoperability mechanisms
- Policy enforcement architectures for distributed systems

The scope of this paper excludes:

- Safety-critical control logic and real-time deterministic control systems
- Failure mode and effects analysis (FMEA) and operational risk modeling
- Physical cybersecurity controls for OT networks
- Product-specific IoT platforms for consumer devices

### IV. DESIGN CONSTRAINTS IN MANUFACTURING IoT ENVIRONMENTS

Manufacturing IoT governance must operate within structural constraints that differ fundamentally from enterprise IT systems. Understanding these constraints is essential to designing governance mechanisms that align with operational reality rather than imposing theoretical models that fail in practice

*A. Asset Longevity and Temporal Asymmetry*

Manufacturing equipment lifecycles extend over 20–30 years, while data platforms evolve on 3–5-year cycles. A PLC installed in 2005 may still be producing data in 2035, interacting with data consumers that did not exist when the equipment was commissioned. Governance models must tolerate long-lived data producers interacting with evolving consumers without requiring retroactive source modifications.

This constraint eliminates governance approaches that depend on source-level standardization or require equipment replacement to achieve semantic consistency. Instead, governance must operate through abstraction layers that translate vendor-specific representations into canonical forms without modifying operational systems

*B. Vendor Hetrogeneity and Semantic Diversity*

PLCs, historians, and gateways expose data using vendor-specific semantics that cannot be normalized in-place. A Rockwell Automation PLC represents process temperature differently than a Siemens S7 controller. Historians from

OSIsoft PI, Honeywell PHD, and GE Proficy use incompatible data models for the same physical measurements.

Attempts to enforce uniform data modeling at the source fail due to vendor lock-in, operational constraints, and the sheer diversity of installed systems [37], [38]. Governance must accommodate semantic diversity through explicit mapping and transformation rather than eliminating it through standardization mandates

*C. Network Segmentation and Reliability Constraints:*

Manufacturing plants operate with strict separation between OT and IT networks to prevent cyber intrusions from affecting production systems. Data governance mechanisms that require bidirectional communication or centralized enforcement across network boundaries violate fundamental operational security principles.

Additionally, plants may operate with intermittent connectivity due to geographic isolation, network infrastructure limitations, or deliberate air-gapping for critical systems. Governance enforcement must function under network partition conditions without blocking local operations [41], [42], [43]

*D. Regulatory and Contractual Obligations*

Regulatory requirements vary by geography (GDPR in Europe, CCPA in California, industry-specific regulations) and product line (medical devices, food safety, environmental monitoring). Contractual obligations with customers, suppliers, and partners impose additional constraints on data sharing and retention.

Policy differentiation is required without fragmenting the data architecture into compliance silos. A single IoT platform may need to enforce different retention policies for personal data versus process telemetry, different access controls for internal analysts versus external auditors, and different cross-border transfer restrictions for EU versus non-EU data

## V. IoT Data Fabric as Governance Context

An IoT data fabric provides a logical abstraction layer that enables standardized governance across distributed data sources without requiring physical centralization. The fabric is not a monolithic platform but a set of shared governance services operating across ingestion, transformation, and consumption layers [1], [5], [35], [36]

The fabric architecture separates concerns across five layers:

- Edge/Plant Systems Layer: OT systems producing raw telemetry without governance responsibilities. These systems operate under plant control and are not modified to accommodate data fabric requirements.

- Ingestion & Canonical Normalization Layer: Gateway services that consume vendor-specific protocols, validate data contracts, and transform raw signals into canonical representations. This layer enforces schema compliance, temporal ordering, and initial quality checks.

- Domain Data Products Layer: Business-capability-aligned data products that aggregate, enrich, and contextualize canonical data for consumption. Domains own semantic interpretation, quality SLAs, and access policies for their products.

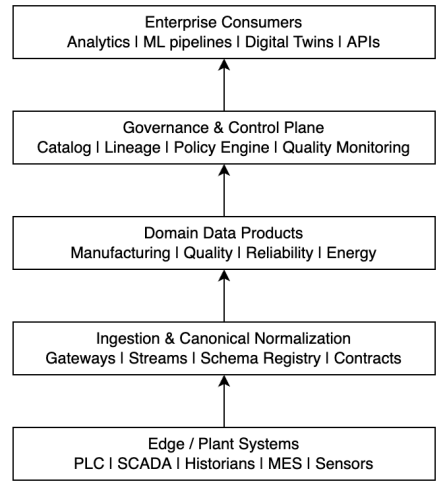

FIG 1. IoT Data Fabric Architecture with Governance Layers

- Governance & Control Plane: Shared services providing metadata catalog, data lineage tracking, policy definition and enforcement, and quality monitoring. This layer operates across all other layers to ensure consistent application of governance.

- Enterprise Consumers Layer: Analytics platforms, machine learning pipelines, digital twin systems, and APIs that consume governed data products. Consumers trust data quality and semantic consistency without direct knowledge of source systems.

Governance applies across all layers through explicit enforcement points at architectural boundaries

## VI. Federated Governance Model

*A. Governance Principles*

The proposed governance model separates policy definition, policy implementation, and policy enforcement. Authority is distributed across organizational layers, while compliance is centralized through shared services. This approach avoids both the bottlenecks of centralized control and the fragmentation of fully autonomous domains.

Three core principles guide the model:
- Subsidiarity: Decisions are made at the lowest level capable of effective implementation. Plants map OT signals to canonical representations. Domains define data products and quality SLAs. Enterprise sets non-negotiable policies.

- Explicit over implicit: All governance decisions are codified as executable artifacts (contracts, policies, schemas) rather than documented as procedures or guidelines. This enables automated enforcement and eliminates interpretation ambiguity.

- Boundaries over perimeters: Governance is enforced at data transition points between architectural layers and

organizational domains, not through perimeter controls around monolithic systems.

### B. Enterprise Governance Layer

The enterprise layer defines foundational policies that apply uniformly across all domains and plants. These policies are intentionally narrow in scope to avoid constraining operational flexibility while maintaining non-negotiable requirements.

Enterprise governance responsibilities include:

- Data Classification Taxonomy: A hierarchical classification scheme (e.g., Public, Internal, Confidential, Restricted) that determines baseline access controls, encryption requirements, and retention policies. Classification is assigned based on data content characteristics, not organizational ownership.

- Access Control Principles: Role-based access control (RBAC) models defining enterprise-wide roles (e.g., Plant Operator, Domain Analyst, Data Steward) and their baseline permissions. Domains may extend but not reduce these permissions.

- Retention and Residency Rules: Regulatory-driven policies specifying minimum and maximum retention periods, geographic residency constraints, and cross-border transfer approvals. These policies are expressed as rules evaluated against asset and data product metadata.

- Canonical Baseline Entities and Events: Core ontology defining fundamental concepts (Asset, Location, Event, Measurement) that all domains must extend rather than redefine. This baseline prevents semantic fragmentation while allowing domain-specific extensions.

Critically, the enterprise layer does not own pipelines, transformations, or operational systems. It provides governance services and enforces policies but does not execute data operations

### C. Domain Governance Layer

Domains represent business capabilities (Manufacturing, Quality, Reliability, Energy) that span multiple plants and product lines. Domain governance councils own the semantic interpretation and operational quality of data products aligned with their capability.

Domain governance responsibilities include:

- Domain-Specific Canonical Extensions: Semantic models extending the enterprise baseline with domain terminology, relationships, and constraints. The Manufacturing domain defines equipment hierarchies, production events, and process parameters. The Quality domain defines defect taxonomies, inspection results, and compliance metrics.

- Data Product Definitions: Explicit specifications of governed datasets published for consumption, including schema, quality SLAs, ownership, and consumption interfaces. A data product is the unit of reuse and accountability in the governance model.

- Schema Evolution Decisions: Policies determining when breaking changes are permitted, how backward compatibility is maintained, and how consumers are notified of changes. These decisions balance innovation velocity with consumer stability.

- Quality Thresholds and SLAs: Quantitative commitments regarding data completeness, accuracy, freshness, and consistency. SLAs are monitored continuously with automated alerting when thresholds are breached.

Domains operate with autonomy within enterprise constraints. They make semantic and operational decisions without enterprise approval but must comply with enterprise policies and publish data products that adhere to canonical baselines.

### D. Plant Governance Layer

Plant governance addresses the operational reality of mapping heterogeneous OT systems to canonical representations while managing local constraints and exceptions.

Plant governance responsibilities include:

- OT Signal Mapping: Explicit mappings from vendor-specific signals (PLC tags, historian points, SCADA variables) to canonical representations. These mappings are maintained as declarative configuration rather than embedded code.

- Local Constraint Management: Handling plant-specific limitations such as network availability, equipment capability, or regulatory requirements that differ from enterprise baselines. Constraints are documented and reconciled with domain expectations.

- Source-Level Remediation: Correcting data quality issues at the source when feasible, including sensor calibration, timestamp synchronization, and duplicate elimination. Issues that cannot be resolved locally are escalated to domain stewards

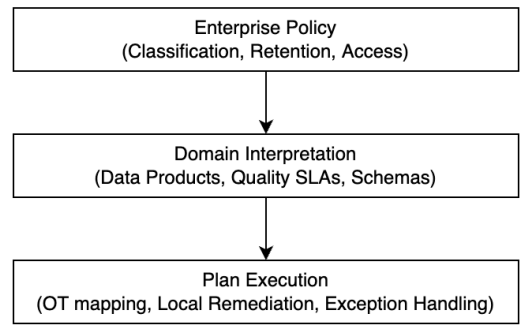

FIG 2. GOVERNANCE AUTHORITY FLOW

Accountability flows upward (plants accountable to domains, domains accountable to enterprise) while operational responsibility flows downward (enterprise does not execute, domains do not operate OT systems)

## VII. IoT Governance Boundaries

Governance is enforced at architectural boundaries where data transitions between trust domains, ownership contexts, or semantic representations. Boundary enforcement enables governance without interfering with real-time operations or requiring modifications to operational systems.

### A. Boundary Identification

Four critical boundaries exist in the reference architecture:

- Boundary 1: OT-to-IT Ingress: Data crosses from operational technology networks into IT infrastructure. This boundary enforces protocol conversion, initial schema validation, and cybersecurity controls.

- Boundary 2: Raw Telemetry to Canonical Representation: Vendor-specific signals are transformed into canonical semantic models. This boundary enforces data contracts, unit normalization, and semantic consistency.

- Boundary 3: Canonical Data to Domain Data Products: Canonical streams are aggregated, enriched, and contextualized into domain-specific data products. This boundary enforces quality SLAs, access policies, and lineage tracking.

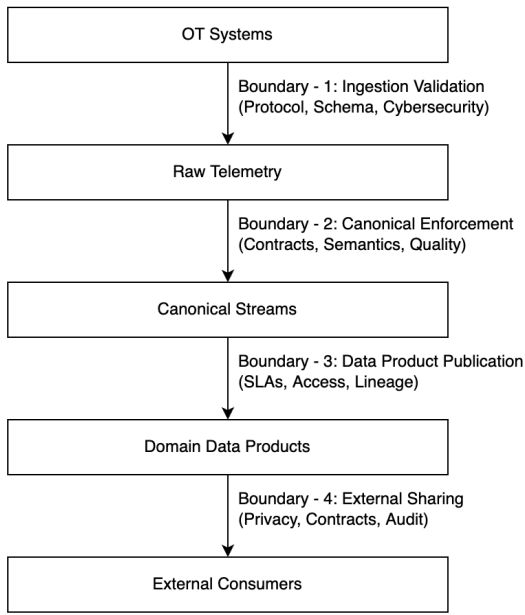

FIG 3. GOVERNANCE BOUNDARY ENFORCEMENT POINTS

- Boundary 4: Internal to External Data Sharing: Data products are exposed to external parties (customers, suppliers, partners, regulators). This boundary enforces contractual obligations, privacy compliance, and audit logging.

### B. Enforcement Mechanisms

Each boundary implements enforcement through a combination of validation, transformation, and policy evaluation:

- Validation: Schema conformance, data type checking, range validation, referential integrity. Validation failures are logged, routed to remediation queues, and do not block compliant data.

- Transformation: Protocol conversion, unit normalization, semantic mapping, timestamp standardization. Transformations are idempotent and preserve lineage metadata.

- Policy Evaluation: Access control decisions, classification assignment, retention application, compliance verification. Policies are evaluated using metadata attributes rather than inspecting payload content.

Enforcement at boundaries avoids the performance penalties and operational risks of inline inspection within real-time data flows

### C. Enforcement Workflow Sequences

The following sequence diagrams illustrate enforcement workflows at critical governance boundaries

*Enforcement Steps:*

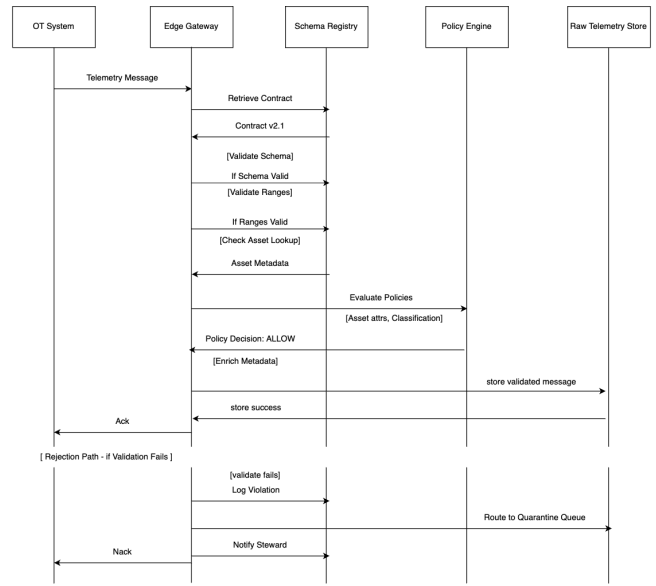

SEQUENCE 1. INGESTION BOUNDARY ENFORCEMENT

1. Edge gateway receives telemetry from the OT system

2. Gateway retrieves the current contract version from the schema registry

3. Schema validation: structure, types, required fields

4. Range validation: min/max bounds, precision checks

5. Referential integrity: asset existence in the registry

6. Policy evaluation: classification, access rules, retention
7. Success path: enrich with metadata, store, acknowledge
8. Failure path: log violation, quarantine, notify steward, negative acknowledgment.

Publication Steps:

1. Domain steward initiates data product publication
2. CI/CD validates contract backward compatibility
3. Pipeline builds a transformation from the canonical to the product schema
4. Quality monitor samples output and measures against SLAs
5. Success path: quality passes, product registered in the catalog, consumers subscribe
6. Failure path: quality fails SLA, deployment blocked, steward alerted.

Access Control Steps:

1. Consumer service requests a data product through the API gateway
2. Gateway extracts user authentication attributes (role, jurisdiction, MFA status)
3. Policy engine evaluates ABAC rules against user and resource attributes
4. Success path: access allowed, conditions applied (masking), data returned, access logged
5. Denial path: policy violation detected, access denied, denial logged with reason

These sequence diagrams demonstrate:

- Ingestion enforcement: Early validation prevents invalid data entry
- Publication enforcement: Quality gates before data product release
- Access enforcement: Runtime policy evaluation with conditional transformations
- Audit trails: Comprehensive logging at all decision points

## VIII. Data Contractors as the Primary Governance Mechanism

Data contracts are the foundational governance artifact in the proposed model [1], [2], [28]. Each IoT dataset or stream published into the fabric is governed by an explicit contract that defines structural, semantic, quality, and operational expectations.

### A. Contract Structure

A data contract consists of six core components:

- Structural Schema: JSON Schema, Apache Avro, or Protocol Buffers definition specifying message structure, field types, required versus optional fields, and nesting relationships. Schema evolution follows semantic versioning (major.minor.patch) with compatibility rules.

- Semantic Meaning and Units: Field-level annotations defining business meaning, engineering units, precision, and relationships to canonical ontology. For example, a field temp_celsius is annotated with unit degC, precision 0.1, and canonical mapping to Measurement.Temperature.

- Temporal Characteristics: Event timestamp semantics (event time versus ingestion time), expected sample rate, allowed timestamp drift, and ordering guarantees. IoT data is inherently temporal, and contracts make temporal semantics explicit.

- Ownership and Stewardship: Identification of the domain owning the contract, the plant or system producing the data, and stewards responsible for resolving quality issues. Ownership enables accountability and remediation workflows.

- Quality Expectations: Quantitative thresholds for completeness (percentage of expected samples received), accuracy (deviation from ground truth), freshness (maximum age), and consistency (referential integrity with related datasets). These expectations form measurable SLAs.

- Versioning and Compatibility Rules: Policies determining when schema changes are breaking versus non-breaking, how consumers are notified, and migration timelines. Compatibility levels (backward, forward, full) are explicitly declared.

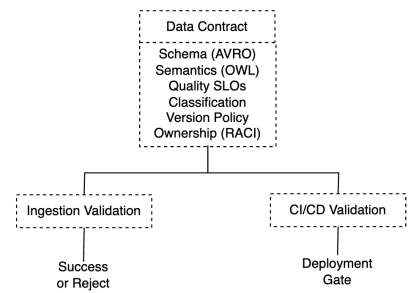

FIG 4. Data Contract Enforcement Architecture

### B. Contract Lifecycle

Data contracts follow a managed lifecycle with automated enforcement at key transitions:

- Definition: A Contract is authored by domain stewards, validated against canonical baselines, and registered in the contract repository with versioning metadata.

- Review: Enterprise and domain governance councils review contracts for policy compliance, semantic

consistency, and operational feasibility. Reviews are asynchronous to avoid blocking operations.

- Deployment: Contracts are deployed to ingestion gateways and validation services via CI/CD pipelines. Deployment includes automated testing against sample data and backward compatibility verification.

- Enforcement: Runtime systems validate incoming data against contracts, rejecting non-compliant messages or routing them to remediation queues. Validation failures trigger alerts to data stewards.

- Evolution: Contract changes are proposed, reviewed, and deployed following semantic versioning rules. Breaking changes require consumer coordination and migration timelines.

- Retirement: Deprecated contracts are marked as end-of-life, consumers are notified and migrated, and enforcement is disabled after a grace period.

*C. Contract-Driven Validation*

Validation occurs at ingestion boundaries before data enters the canonical layer. Validation failures are classified by severity:

- Critical: Missing required fields, type mismatches, referential integrity violations. These failures block ingestion and trigger immediate alerts.

- Warning: Range exceedances, unexpected null values, minor schema deviations. These failures are logged but do not block ingestion.

- Informational: Deprecation notices, quality threshold approaches, schema evolution notifications. These are tracked for trend analysis.

Validation metrics (failure rates, failure types, time-to-remediation) are continuously monitored and published to governance dashboards

## IX. ASSET AND IDENTIFY CENTRIC GOVERNANCE

IoT data governance must be asset-centric rather than dataset-centric. Every signal carries implicit context about the physical asset producing it, including location, operational state, and lifecycle stage. Making this context explicitly enables fine-grained governance without duplicating pipelines or schemas.

*A. Asset Identity Model*

The governance model relies on a hierarchical asset identity model that aligns with manufacturing organizational structures:

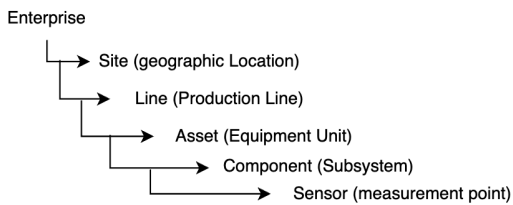

FIG 5. HIERARCHY ASSET IDENTITY MODEL

Each level in the hierarchy carries attributes used for policy evaluation:

- Enterprise: Regulatory jurisdiction, data residency requirements, corporate policies

- Site: Geographic coordinates, local regulations, network infrastructure, operational hours

- Line: Product family, production schedule, quality requirements, energy constraints

- Asset: Manufacturer, model, installation date, maintenance schedule, criticality tier

- Component: Part number, firmware version, expected lifespan, calibration status

- Sensor: Measurement type, units, sample rate, accuracy specification, last calibration

Asset identities are globally unique and stable across the asset lifecycle. When an asset is relocated, its identity persists with updated location attributes. When a component is replaced, the new component inherits the identity with updated installation metadata.

*B. Identity-Driven Policy Enforcement*

Access, retention, and sharing policies are expressed using asset metadata rather than static dataset identifiers. This approach provides several advantages:

- Fine-Grained Control: Policies can target specific asset classes, locations, or lifecycle states without enumerating all affected datasets. For example, "retain quality inspection data for assets in EU sites for 10 years" applies automatically to all relevant data products.

- Lifecycle Awareness: Policies adapt to asset lifecycle transitions without manual updates. When an asset moves from production to decommissioning, retention policies automatically adjust.

- Attribute-Based Access Control (ABAC): Access decisions evaluate requestor attributes (role, clearance, jurisdiction) against resource attributes (asset classification, location, sensitivity) using policy rules rather than static access control lists.

- Temporal Governance: Policies reference asset temporal states (commissioning, operation, maintenance, decommissioning) to apply context-appropriate governance. Data produced during commissioning may have different retention requirements than operational data

*C. Device Identity and Attestation*

For connected products and edge devices, device identity management extends the asset model with cryptographic assurance [3], [30]. Each device possesses a unique Device Attestation Certificate (DAC) provisioned during manufacturing. Device identity enables:

- Authenticated Ingestion: Ingestion gateways verify device identity before accepting data, preventing spoofing and injection attacks.

- Device Lifecycle Tracking: Governance policies reference device lifecycle states (provisioned, commissioned, active, suspended, revoked) to control data handling.

- Secure Credential Management: Device credentials are rotated automatically based on governance policies without manual intervention.

- Audit Trails: All data carries device identity metadata, enabling comprehensive audit trails for compliance verification

## X. Policy-as-code-Enforcement

Governance policies are implemented as executable rules evaluated by shared services rather than documented procedures executed manually. Policy-as-code enables consistent, auditable, and automated governance at scale [4], [23].

### A. Policy Categories

Four policy categories address distinct governance concerns:

- Access Policies: Define who can access which data products under what conditions. Policies specify roles, attribute requirements, approval workflows, and audit logging.

- Security Policies: Enforce encryption, network controls, and cybersecurity requirements. Policies reference data classification and asset criticality to determine required controls.

- Compliance Policies: Implement regulatory requirements including retention minimums and maximums, cross-border transfer restrictions, and data subject rights. Policies are jurisdiction-aware and asset-location-sensitive.

- Quality Policies: Define completeness, accuracy, freshness, and consistency thresholds with automated monitoring and alerting. Quality policies are associated with data product SLAs

### B. Policy Categories

Policies are expressed in a domain-specific language (DSL) that balances expressiveness with operational clarity. The DSL supports:

- Attribute-Based Rules: Conditions referencing asset attributes, user attributes, temporal context, and data characteristics.

- Temporal Logic: Time-based conditions including retention periods, access windows, and lifecycle transitions.

- Composability: Policy fragments that can be combined, extended, and overridden following inheritance hierarchies.

- Versioning: Policy evolution with effective dates, deprecation notices, and migration paths.

### C. Enforcement Architecture

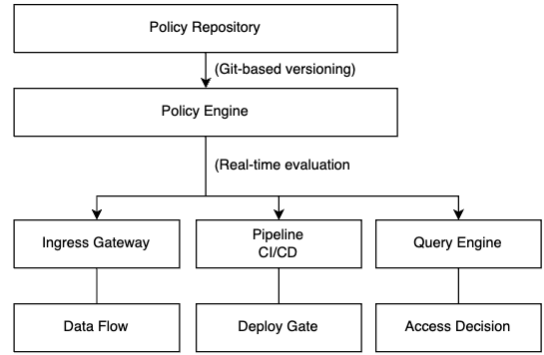

FIG 6. Policy-as-Code Enforcement Architecture

The policy engine evaluates policies continuously at enforcement points:

- Ingress Gateway: Access and security policies evaluated before data enters the fabric. Non-compliant data is rejected or quarantined.

- Pipeline CI/CD: Quality and compliance policies evaluated before deploying transformations or data products. Policy violations block deployment.

- Query Engine: Access policies evaluated at query time to filter results based on requestor attributes and resource sensitivity.

### D. Policy Observability

Policy enforcement generates observability data including:

- Evaluation Metrics: Policy evaluation latency, decision distribution (allow/deny/escalate), cache hit rates

- Violation Events: Policy violation details, affected resources, remediation status

- Compliance Reports: Aggregated policy adherence across domains, plants, and policy categories

- Audit Logs: Immutable records of policy decisions, including requestor identity, resource accessed, decision outcome, and evaluation timestamp

Observability data feeds governance dashboards and regulatory reporting systems

## XI. Schema Governance and Evolution

Schema governance addresses the challenge of evolving data structures without breaking consumer systems. IoT platforms must balance innovation velocity (enabling rapid schema evolution) with consumer stability (preventing unexpected breakage).

## A. Schema Registry

A centralized schema registry serves as the authoritative source for all schemas used across the data fabric [2], [5], [7]. The registry provides:

- Version Control: All schema versions are retained with metadata describing changes, compatibility levels, and deprecation status.

- Compatibility Checking: Automated validation of proposed schema changes against compatibility rules (backward, forward, full, none).

- Lineage Tracking: Visibility into which data products and consumers use each schema version, enabling impact analysis for changes.

- Search and Discovery: Metadata-driven search enabling consumers to find schemas by domain, asset type, or semantic concept.

## B. Schema Evolution Rules

Schema evolution follows semantic versioning principles adapted for streaming data:

- Patch Version (1.0.0 → 1.0.1): Documentation updates, metadata corrections, no structural changes. Fully compatible.

- Minor Version (1.0.0 → 1.1.0): Additive changes only (new optional fields, new message types). Backward compatible (old producers, new consumers).

- Major Version (1.0.0 → 2.0.0): Breaking changes (removing fields, changing types, reordering fields). Not backward compatible, requires consumer migration.

Domains establish migration timelines for major version changes, typically 3–6 months to allow consumer updates.

## C. Canonical Model Governance

The canonical data model serves as the integration layer across heterogeneous source systems. Canonical model governance ensures semantic consistency without constraining domain innovation.

- Baseline Ontology: Core concepts (Asset, Event, Measurement, Location) defined by enterprise governance. Domains extend but do not redefine baseline concepts.

- Extension Mechanisms: Domains introduce new entity types, properties, and relationships through formal extension processes. Extensions are reviewed for semantic conflicts before approval.

- Versioning Strategy: Canonical model evolves through versioned releases coordinated across domains. Breaking changes require cross-domain coordination.

- Mapping Specifications: Explicit mappings from vendor-specific source schemas to canonical representations. Mappings are maintained as declarative configuration and version-controlled alongside schemas.

## XII. QUALITY GOVERNANCE AND OBSERVABILITY

Quality governance addresses the challenge of maintaining trust in IoT data despite equipment malfunctions, network issues, and configuration errors. Quality is not achieved through inspection alone but through continuous monitoring, automated remediation, and clear accountability.

### A. Quality Dimensions

IoT data quality is assessed across five dimensions [6]:

- Completeness: Percentage of expected data points received relative to asset sample rate and operational schedule. Completeness degradation indicates sensor failure, network loss, or configuration drift.

- Accuracy: Deviation between measured values and ground truth or expected ranges. Accuracy issues indicate sensor calibration errors or equipment malfunction.

- Freshness: Age of data relative to event time. Freshness degradation indicates network latency, processing backlog, or timestamp synchronization issues.

- Consistency: Referential integrity across related datasets. Consistency violations indicate incomplete data product publication or lineage breakage.

- Validity: Conformance to schema contracts, unit specifications, and value constraints. Validity failures indicate source configuration errors or contract drift.

### B. Quality Monitoring Architecture

Quality metrics are collected continuously through sampling (not exhaustive inspection to avoid performance impact). Metrics are aggregated by asset, data product, and domain with configurable time windows.

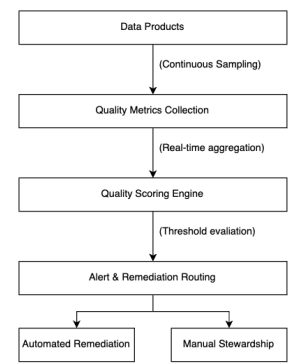

FIG 7. QUALITY GOVERNANCE FRAMEWORK ARCHITECTURE

Quality scores combine weighted dimensions into composite indicators. Scores below SLA thresholds trigger automated alerts routed to responsible stewards based on RACI (Responsible, Accountable, Consulted, Informed) models defined in data contracts

## C. Remediation Workflows

Quality issues are routed to remediation workflows based on failure classification:

- Automated Remediation: Interpolation for missing samples, outlier detection and suppression, timestamp correction, duplicate elimination. Automated remediation is applied transparently with lineage metadata recording the intervention.

- Semi-Automated Remediation: Rules-based routing to stewards with suggested actions. For example, sustained completeness degradation triggers investigation workflow with recommended diagnostics.

- Manual Stewardship: Complex quality issues requiring domain expertise, source investigation, or cross-domain coordination. Stewardship workflows track issue status, root cause analysis, and resolution timelines.

Remediation effectiveness is measured through mean-time-to-detect (MTTD) and mean-time-to-resolve (MTTR) metrics published to governance dashboards

## XIII. ACCESS GOVERNANCE AND PRIVACY

Access governance balances data democratization (enabling broad access for analytics and innovation) with privacy protection and regulatory compliance. The governance model implements attribute-based access control with dynamic policy evaluation.

### A. Access Control Model

Access decisions evaluate three attribute sets:

- Subject Attributes: Requestor role, clearance level, organizational affiliation, geographic location, authentication assurance level.

- Resource Attributes: Data classification, asset location, sensitivity level, regulatory constraints, lifecycle stage.

- Environmental Attributes: Request timestamp, access channel, purpose of use, data residency requirements.

Policies express conditional logic combining attribute sets [26]. For example: "Allow access to Confidential manufacturing data for users with Analyst role AND located in same jurisdiction as asset AND authenticated with MFA."

### B. Privacy-Preserving Techniques

For data products containing personally identifiable information (PII) or sensitive operational details, privacy-preserving techniques are applied:

- Tokenization: Replacing identifiers with reversible tokens, enabling analysis without exposing source identities. Token mapping is maintained in secured registries with access controls.

- Aggregation: Publishing aggregated statistics rather than raw measurements where individual-level data is unnecessary. Aggregation levels are policy-driven based on data classification.

- Differential Privacy: Adding calibrated noise to query results to prevent re-identification while maintaining statistical utility. Privacy budgets are tracked per data product.

- Purpose Limitation: Requiring explicit purpose declarations for data access, logging purposes for audit, and restricting repurposing without additional approval.

### C. Cross-Border data Governance

Global manufacturing platforms must navigate varying data sovereignty requirements. Cross-border governance is implemented through [21], [24]:

- Data Residency Policies: Specifying which asset data must remain in-region based on regulatory requirements (e.g., GDPR, China Cybersecurity Law).

- Transfer Mechanisms: Implementing standard contractual clauses, binding corporate rules, or explicit consents for cross-border transfers where permitted.

- Geographic Partitioning: Physically partitioning data storage and processing by jurisdiction to simplify compliance verification.

- Metadata Synchronization: Replicating metadata globally while restricting payload access by geography, enabling discovery without violating residency constraints.

## XIV. GOVERNANCE OUTCOMES AND DESIGN OBJECTIVES

The federated governance model is designed to produce measurable operational outcomes that justify investment and guide continuous improvement. This section defines expected outcomes and success metrics for governance implementation.

### A. Quantitative Outcome Targets

- Schema Breakage Reduction: Percentage reduction in downstream pipeline failures due to unexpected schema changes. The model's contract-based approach with CI/CD validation and semantic versioning is designed to prevent breaking changes from reaching production. Target: 80% reduction in schema-related incidents within 12 months of contract enforcement deployment.

- Data Reuse Increase: Growth in the number of distinct consumers per data product, indicating improved trust and discoverability through explicit contracts and quality SLAs. The model's data product approach with discoverable contracts should enable cross-domain reuse. Target: 3× increase in cross-domain data product consumers within 18 months.

- Quality SLA Adherence: Percentage of data products meeting defined quality thresholds consistently through continuous monitoring and automated remediation. The model's quality governance with threshold-based alerting and remediation workflows should improve adherence. Target: 95% of data products meeting quality SLAs consistently.

- Incident Resolution Velocity: Reduction in mean-time-to-resolve (MTTR) for data quality incidents through

improved lineage tracking and ownership clarity. The model's explicit ownership in contracts and automated lineage should accelerate remediation. Target: 50% MTTR reduction within 12 months.

- Compliance Verification Efficiency: Reduction in effort required to demonstrate regulatory compliance through automated audit trails and policy-as-code evidence. The model's comprehensive logging and policy enforcement should streamline compliance. Target: 70% reduction in compliance verification effort.

These targets represent design objectives based on governance theory and the model's architectural features. Actual outcomes would require empirical validation through controlled deployment.

*B. Qualitative Outcome Objectives*

- Organizational Clarity: Unambiguous assignment of ownership and stewardship responsibilities through explicit RACI models in contracts and federated governance structures. This should eliminate confusion during incidents and enable faster escalation.

- Innovation Acceleration: Reduced friction in accessing and trusting IoT data for analytics through discoverable contracts and quality guarantees, enabling faster experimentation and model development without extensive data quality investigation.

- Regulatory Confidence: Demonstrable compliance posture through comprehensive audit trails, policy-as-code evidence, and automated enforcement, reducing compliance risk and audit preparation effort.

- Operational Resilience: Faster recovery from data quality incidents through automated remediation workflows, clear escalation paths, and explicit quality SLAs with steward notification.

*C. Governance Maturity Progression*

Organizations implementing the governance model would progress through maturity stages:

- Stage 1 - Reactive (Initial State): Ad-hoc governance, manual processes, unclear ownership. Governance is invoked only during incidents. No contracts, no automated validation, no explicit policies.

- Stage 2 - Defined: Documented policies and procedures, assigned roles, basic tooling. Governance is understood but not consistently applied. Contracts may be defined, but enforcement is manual or partial.

- Stage 3 - Managed: Data contracts deployed, automated validation at boundaries, quality monitoring operational. Governance is embedded in operational workflows. Policies are defined but may not be fully automated.

- Stage 4 - Optimized: Policy-as-code enforcement, continuous improvement, predictive quality management. Governance is transparent, self-service, and continuously refined based on metrics. Full automation of routine decisions with human oversight for exceptions.

Maturity advancement would be measured through governance metrics, with targeted improvement initiatives addressing capability gaps at each stage.

*D. Success Measurement Framework*

Implementation success should be evaluated through:

- Process Metrics: - Contract coverage: percentage of data streams governed by contracts - Policy enforcement rate: percentage of governance decisions automated - Validation failure rate: trending over time (expect initial increase as violations are detected, then decrease as sources are corrected) - Quality SLA compliance rate: percentage of time products meet thresholds

- Operational Metrics: - Time-to-publish: duration from contract definition to data product availability - Time-to-resolution: incident detection to remediation completion - Cross-domain reuse ratio: consumers per data product - Schema evolution velocity: contract updates per quarter with compatibility preservation

- Business Metrics: - Compliance audit preparation effort: hours required - Data-driven decision latency: time from question to trusted answer - Analytics project acceleration: time from concept to production model - Incident cost reduction: prevented failures and faster remediation

- Organizational Metrics: - Governance council effectiveness: decisions per meeting, time to decision - Steward engagement: active participation in quality remediation - Domain autonomy: percentage of decisions made without enterprise escalation - Plant capability: successful mapping maintenance without external support

These metrics provide a comprehensive framework for evaluating the effectiveness of governance implementation, though actual measurement would require production deployment.

XV. IMPLEMENTATION CONSIDERATIONS

This section describes implementation considerations for deploying the federated governance model in global manufacturing environments. The considerations are derived from analysis of manufacturing IoT platform requirements, architectural constraints, and organizational structures typical of medium-to-large industrial enterprises.

*A. Deployment Scenarios and Requirements*

The governance model is designed to address three representative deployment scenarios that capture the range of challenges in manufacturing IoT governance.

*Scenario 1: Connected Vehicle Platform*

A global automotive manufacturer operates a connected vehicle platform serving millions of vehicles across multiple continents. The platform ingests telemetry from vehicle subsystems (powertrain, battery management, infotainment, advanced driver assistance systems) at high event rates during peak traffic periods.

- Governance Requirements: - Heterogeneous OEM and tier-1 supplier data formats requiring semantic normalization - Regional data residency requirements (GDPR, China Cybersecurity Law) necessitating geographic partitioning - Real-time quality monitoring without introducing latency to time-critical diagnostics - Contract evolution coordination across distributed engineering teams

- Architecture Considerations: - Cloud-based ingestion infrastructure (AWS IoT Core, Azure IoT Hub, or Google Cloud IoT Core) - Event streaming platform (Apache Kafka, Amazon Kinesis, Azure Event Hubs) - Schema registry for contract management (Apache Schema Registry, AWS Glue, Azure Schema Registry) - Policy engine for access control (Open Policy Agent, AWS Cedar) - Multiple domain data products organized by business capability (battery health, predictive maintenance, crash detection)

*Scenario 2: Manufacturing Production IoT Platform*

A multinational manufacturer operates production facilities producing industrial equipment. The platform monitors production assets (CNC machines, injection molding equipment, assembly robots, environmental chambers) across geographically distributed plants.

- Governance Requirements: - Legacy equipment with proprietary protocols requiring custom gateway development - Plant network segmentation preventing direct cloud connectivity for some facilities - Varying data maturity across plants (some with historians, others with manual collection) - Quality data linking production telemetry to compliance certifications for regulatory reporting

- Architecture Considerations: - Edge gateway deployments with local processing capability - Hybrid cloud-edge architecture supporting disconnected operation - Metadata cataloging with lineage tracking (Apache Atlas, AWS Glue Data Catalog, Azure Purview) - Custom policy enforcement at edge gateways for network-segmented environments - Canonical measurement types mapped across manufacturing domains (machining, assembly, testing, energy, quality)

*Scenario 3: Connected Products Platform*

A manufacturer operates a connected products platform for residential and commercial equipment enabling demand response, predictive maintenance, and energy optimization through cloud connectivity.

- Governance Requirements: - Device firmware diversity across product generations spanning multiple years - Customer privacy requirements for residential installations - Third-party data sharing requiring governed consent mechanisms - Over-the-air firmware updates requiring contract compatibility verification before deployment

- Architecture Considerations: - Device identity management with cryptographic attestation (Device Attestation Certificates) - Message queuing infrastructure (Google Cloud Pub/Sub, Amazon SQS, RabbitMQ) - Device state management (Firestore, DynamoDB, Cosmos DB) - Policy enforcement leveraging device attestation and lifecycle states - Data contracts covering device telemetry, user interactions, utility signals, maintenance events

*B. Design Patterns and Observations*

Analysis of these deployment scenarios reveals consistent design patterns and considerations that inform governance model implementation.

*Pattern 1: Contract-Driven Semantic Alignment*

Explicit contract definition addresses latent semantic ambiguity that informal documentation cannot resolve. In connected vehicle scenarios, defining contracts for measurements like "vehicle speed" requires disambiguating multiple interpretations: wheel speed from ABS sensors, GPS-derived ground speed, or speedometer display value. Each interpretation has different precision, update rates, and availability conditions.

Design Implication: Contract authoring should surface semantic conflicts early through formal definition requirements. Contract templates and authoring tools reduce friction while enforcing completeness.

*Pattern 2: Boundary Enforcement Exposes Coupling*

Automated enforcement at architectural boundaries reveals implicit dependencies between supposedly independent data streams. Manufacturing scenarios demonstrate that measurements from downstream equipment often have timing dependencies on upstream measurements. When network latency reorders events, downstream validation may fail.

Design Implication: Validation logic must account for temporal dependencies and event ordering guarantees. Contract specifications should explicitly declare ordering requirements and correlation constraints.

*Pattern 3: Federated Governance Requires Organizational Alignment*

Technical governance mechanisms alone are insufficient without organizational structures supporting federated decision-making. Manufacturing organizations with strong functional silos between IT and OT may struggle to implement domain governance councils that span these boundaries.

Design Implication: Governance implementation requires parallel organizational change. Plant engineering teams must understand their role as owners of OT-to-canonical mappings, not underlying equipment behavior. Domain councils require executive sponsorship and clear decision rights.

*Pattern 4: Automation Scales, Manual Processes Do Not*

Governance mechanisms requiring manual intervention for routine decisions create bottlenecks that organizations route around. Access control policies requiring manual approval for each request quickly accumulate backlogs, leading to unauthorized workarounds.

Design Implication: Automate all routine governance decisions with explicit approvals only for exceptions. Policy evaluation must operate at system timescales (milliseconds for ingestion, hours for policy changes) not human timescales (days for reviews).

*Pattern 5: Quality Validation Requires Domain Expertise*

Generic quality metrics (completeness, freshness, validity) detect obvious failures but miss domain-specific anomalies. Manufacturing temperature sensors may pass baseline quality checks while providing incorrect data due to calibration drift. Domain experts recognize patterns (variance too low for expected process behavior) that automated checks miss.

Design Implication: Quality governance combines automated baseline checks with domain-specific validation rules. Domain councils must define expected patterns (acceptable ranges, temporal correlations, cross-signal consistency) that supplement generic metrics.

### C. Implementation Strategy

Successful governance implementation follows a phased approach that demonstrates value incrementally while building organizational capability.

*Phase 1: Pilot Domain Selection*

Select one domain with manageable scope (8-12 contracts) to demonstrate feasibility and value. Pilot domains should have: - Clear ownership structure with identified stewards - Downstream consumers willing to collaborate on contract definition - Existing quality pain points that governance can address - Technical capability to implement contracts without extensive training

Avoid attempting enterprise-wide rollout simultaneously across all domains. The coordination overhead and organizational change requirements overwhelm teams and delay value realization.

*Phase 2: Contract Authoring Infrastructure*

Invest in contract authoring tools that reduce friction for domain experts. Raw schema languages (JSON Schema, Avro) are too verbose and error-prone for non-specialists. Simplified authoring interfaces include: - Web-based forms for common contract patterns - Template libraries for standard measurement types - Validation feedback integrated into authoring workflows - Version control integration for contract lifecycle management

*Phase 3: Automated Enforcement Deployment*

Implement enforcement at architectural boundaries with automated validation and policy evaluation. Initial deployment should focus on: - Ingestion validation preventing invalid data entry - Schema compatibility checking in CI/CD pipelines - Quality monitoring with alerting to stewards - Access policy evaluation at query boundaries

Manual review should be reserved for exceptions and policy violations requiring human judgment.

*Phase 4: Measurement and Iteration*

Establish metrics for governance effectiveness: - Contract compliance rates (percentage of messages passing validation) - Quality SLA adherence (percentage of time domains meet quality thresholds) - Incident resolution velocity (time from quality degradation to remediation) - Cross-domain reuse (number of distinct consumers per data product)

Use metrics to drive continuous improvement: identify validation rules causing false positives, quality thresholds requiring adjustment, and policy gaps requiring new rules.

*Phase 5: Incremental Expansion*

Expand governance to additional domains based on demonstrated value and organizational readiness. Expansion should be demand-driven (domains requesting adoption) rather than top-down mandates.

Typical implementation timeline: 18-24 months from pilot initiation to enterprise coverage across major domains.

### D. Technology Enablers

Implementing the governance model requires specific technology capabilities:

- Schema Registry: Apache Schema Registry, AWS Glue Schema Registry, or Azure Schema Registry supporting Avro, Protobuf, and JSON Schema with compatibility checking.

- Policy Engine: Open Policy Agent (OPA), AWS Cedar, or custom rule engines capable of evaluating policies at ingestion, deployment, and query boundaries.

- Metadata Catalog: Apache Atlas, AWS Glue Data Catalog, or Azure Purview providing searchable metadata with lineage tracking.

- Data Quality Framework: Great Expectations, Apache Griffin, or Datadog Data Observability for quality metric collection and threshold evaluation.

- Identity and Access Management: IAM systems supporting ABAC with integration to asset metadata stores for context-aware access decisions.

### E. Organizational Enablers

Technology alone is insufficient. Organizational capabilities are equally critical:

- Governance Councils: Cross-functional bodies including domain stewards, data engineers, compliance officers, and operational representatives making governance decisions.

- Role Definitions: Clear RACI matrices defining who is Responsible, Accountable, Consulted, and Informed for governance activities.

- Training Programs: Capability development for data stewards, domain analysts, and plant engineers on governance concepts, tools, and workflows.
- Incentive Alignment: Performance metrics and reward systems recognizing governance contributions, including data product quality, contract compliance, and cross-domain reuse.

*F. Migration Strategy*

Existing IoT platforms require phased migration rather than wholesale replacement [27]:

- Phase 1 - Pilot Domain: Select one domain with a manageable scope, implement contracts and quality monitoring, and demonstrate value through metrics.
- Phase 2 - Expand Coverage: Incrementally onboard additional domains, reusing governance services and tooling, refining processes based on lessons learned.
- Phase 3 - Policy Automation: Transition from manual policy enforcement to policy-as-code, automate validation and remediation.
- Phase 4 - Continuous Improvement: Establish feedback loops, track governance metrics, and invest in capability maturity advancement.

Migration timelines typically span 18–24 months for enterprise-scale implementations.

## XVI. Future Directions

*A. AI-Driven Governance*

Machine learning can enhance governance through:

- Anomaly Detection: Identifying quality degradation patterns not captured by static rules, enabling proactive remediation.
- Schema Inference: Automatically generating contract drafts from observed data patterns, reducing manual effort.
- Policy Optimization: Analyzing policy violation patterns to recommend policy adjustments balancing control and operational flexibility.

*B. Federated Learning for Quality*

Federated learning enables quality model training across plants without centralizing sensitive operational data, respecting data residency constraints while improving quality prediction.

*C. Blockchain for Audit Trails*

Distributed ledger technologies provide immutable audit trails for governance decisions, enhancing regulatory compliance verification and trust in multi-party collaborations [29].

## XVII. Limitations and Future Work

This work presents a systems architecture and design framework for IoT data governance. As a design contribution, several important limitations and areas for future research must be acknowledged.

*A. Lack of Empirical Validation*

The governance model presented is a design framework grounded in the analysis of manufacturing IoT requirements and constraints. It has not been validated through controlled empirical studies or production deployments. Future work must address this gap through structured validation across diverse organizational contexts. Required Empirical Work:

- Controlled Experiments: Comparing federated governance against centralized and fully autonomous approaches using randomized controlled trials. Dependent variables should include schema breakage rates, data quality metrics, cross-domain reuse, and incident resolution times.
- Case Study Research: Longitudinal studies documenting governance adoption in multiple manufacturing organizations with varying maturity levels, organizational structures, and technical capabilities. Case studies should capture adoption challenges, patterns of organizational resistance, and quantitative governance metrics using standardized measurement protocols.
- Generalization Testing: Validation beyond automotive and HVAC manufacturing to other sectors, including pharmaceuticals, food processing, discrete electronics, and chemical processing. Each sector presents unique regulatory requirements, equipment characteristics, and organizational structures that may require model adaptation.
- Scalability Validation: Testing at varying scales from small manufacturers (single plant, hundreds of assets) to global enterprises (dozens of plants, tens of thousands of assets) to identify scalability limitations and required model modifications.

Without empirical validation, claims about governance effectiveness, adoption feasibility, and organizational impact remain hypothetical.

*B. Formal Verification*

The governance model's correctness properties—particularly policy conflict detection, completeness of boundary enforcement, and contract compatibility verification—have not been formally verified. While the model specifies validation mechanisms and enforcement points, formal proofs demonstrating that governance invariants hold under all operational conditions are future work. Verification Challenges:

- Policy Conflict Resolution: Demonstrating that federated policy composition across enterprise, domain, and plant layers produces deterministic enforcement decisions without contradictions. Policies may be defined independently at each layer, potentially creating conflicts when composed.
- Boundary Completeness: Proving that all data transitions between architectural layers pass through

governance enforcement points without bypass paths. Formal verification should demonstrate that the architecture prevents ungoverned data flows.

- Contract Evolution Safety: Verifying that schema evolution following semantic versioning rules prevents breaking changes from reaching consumers without migration coordination. Proof should cover compatibility checking algorithms and consumer notification mechanisms.

- Liveness Properties: Demonstrating that governance enforcement does not introduce deadlocks or livelocks that block operational data flows. Policies must be evaluated efficiently enough to meet real-time requirements without causing cascading delays.

Formal methods (temporal logic, model checking, theorem proving) could strengthen confidence in governance correctness. Integration with policy verification frameworks (OPA testing, AWS Cedar policy validation) would provide automated verification capabilities.

*C. Organizational Structure Assumptions*

The federated governance model assumes specific organizational structures and capabilities that may not exist in all manufacturing enterprises:

- Domain Capability Requirements: The model assumes domains can assign dedicated data stewards, define data products, and own semantic interpretation. Smaller organizations or those with less mature data capabilities may lack the personnel or expertise to operate domain governance councils effectively.

- Plant Technical Capacity: The model assumes plants have sufficient technical capability to maintain OT signal mappings, execute local remediation, and participate in governance escalations. Plants with limited engineering staffing or vendor-managed operational systems may struggle to fulfill these responsibilities.

- Cross-Functional Collaboration: Federated governance requires ongoing collaboration between enterprise governance (setting policies), domain councils (interpreting requirements), and plant operations (executing mappings). Organizations with strong functional silos or adversarial relationships between IT and OT may face adoption challenges not addressed by the technical architecture.

- Change Management Capacity: Implementing the governance model requires organizational change - new roles, new processes, new tools. Organizations undergoing simultaneous transformations (mergers, divestitures, major system replacements) may lack capacity for governance transformation.

- Executive Sponsorship: Domain governance councils require executive sponsorship to make cross-functional decisions and allocate resources. Without executive support, councils may lack authority to enforce accountability.

The model is designed for medium-to-large manufacturing enterprises with distributed operations and sufficient organizational maturity to support federated structures. Adaptation for small manufacturers (single plant, centralized decision-making) or highly centralized organizations requires modification of the governance layers and accountability structures.

*D. Technology Stack Dependencies*

While the model emphasizes provider-agnostic mechanisms, practical implementations depend on specific technology choices with associated limitations:

- Schema Registry Capabilities: Compatibility checking, versioning semantics, and performance characteristics vary across implementations (Apache Schema Registry, AWS Glue, Azure Schema Registry). The model assumes registry capabilities that may not be universally available or may require custom development.

- Policy Engine Expressiveness: Policy-as-code enforcement assumes engines (OPA, Cedar) capable of evaluating complex attribute-based policies with acceptable latency. Limitations in policy language expressiveness or evaluation performance may constrain governance sophistication.

- Metadata Catalog Integration: Lineage tracking, search capabilities, and asset metadata management depend on catalog implementations (Apache Atlas, AWS Glue Data Catalog, Azure Purview). Integration challenges across heterogeneous catalogs may limit governance visibility.

- Network and Compute Resources: Boundary enforcement requires sufficient network bandwidth and compute resources at enforcement points. Resource-constrained environments (edge gateways with limited CPU/memory) may struggle to execute complex validation and policy evaluation.

Future work should document minimum technology requirements, provide reference implementations demonstrating interoperability across platforms, and identify graceful degradation strategies when ideal technology capabilities are unavailable.

*E. Security and Privacy Analysis*

The governance model addresses access control, data classification, and privacy-preserving techniques but does not comprehensively address all security and privacy requirements for manufacturing IoT:

- Threat Modeling: Systematic threat analysis of governance enforcement points, policy repositories, and metadata catalogs has not been conducted. Adversarial attacks targeting governance infrastructure could undermine enforcement without detection.

- Privacy Budget Management: Differential privacy techniques are mentioned but detailed privacy budget allocation, composition theorems, and utility-privacy tradeoffs are not specified. Implementation would require careful analysis to prevent privacy leakage through repeated queries.

- Cryptographic Enforcement: While device identity leverages cryptographic attestation, end-to-end encryption of data products and cryptographic policy enforcement are not addressed. Implementations may require cryptographic mechanisms beyond device attestation.

- Insider Threats: The model assumes administrators and data stewards act in accordance with policies. Insider threats (malicious administrators, compromised credentials) are not explicitly addressed through technical controls.

Comprehensive security and privacy analysis, including threat modeling, cryptographic protocol design, privacy impact assessment, and insider threat mitigation, is planned for future work.

*F. Economic Analysis*

The governance model does not include economic analysis of implementation costs, operational overhead, or return on investment:

- Implementation Costs: Capital expenditure for technology infrastructure, development effort for custom integrations, organizational change management, and training programs.

- Operational Costs: Ongoing personnel costs (data stewards, governance councils, policy administrators), infrastructure costs (compute, storage, network), and tool licensing.

- Benefit Quantification: Methods for measuring and attributing improvements in data quality, reduced incidents, faster consumer onboarding, and regulatory compliance efficiency to governance mechanisms versus other concurrent changes.

- Cost-Benefit Models: Decision frameworks helping organizations determine when federated governance investment is justified versus simpler centralized or autonomous approaches.

Future work should develop economic models enabling organizations to evaluate governance investment decisions and optimize resource allocation across governance activities.

## XVIII. CONCLUSION

This paper presented a federated governance model for IoT data in global manufacturing environments. The model addresses the unique challenges of distributed operational ownership, heterogeneous source systems, and continuous change through contract-driven interoperability, policy-as-code enforcement, and asset-centric accountability. Key contributions include:

- Federated governance architecture separating policy definition, implementation, and enforcement across organizational layers while maintaining centralized compliance verification

- Boundary-based enforcement operating at architectural transitions rather than within operational systems, respecting manufacturing's operational constraints

- Data contracts as governance mechanism ensuring semantic consistency and quality through explicit, versioned specifications

- Asset-centric policy application enabling fine-grained governance without schema fragmentation through hierarchical asset identity models

- Policy-as-code framework with formal grammar enabling automated, auditable governance at scale

- Comprehensive positioning against existing approaches (data mesh, IIRA, cloud providers, data spaces) demonstrating specific contributions to manufacturing IoT governance

- Complete technical specifications including BNF grammar, contract schemas, and enforcement workflows enabling implementation

This work contributes a systems architecture and design framework grounded in analysis of manufacturing IoT requirements and constraints. The model is designed for medium-to-large manufacturing enterprises with distributed operations requiring governance that balances operational autonomy with centralized compliance.

As a design contribution, this work provides the architectural foundation and technical specifications required for implementation. However, as acknowledged in Section XVII, empirical validation through controlled studies, production deployments, and longitudinal case research remains essential future work. Specific areas requiring validation include:

- Governance effectiveness metrics across diverse organizational contexts

- Adoption feasibility and organizational change requirements

- Scalability characteristics from small manufacturers to global enterprises

- Formal verification of policy conflict resolution and boundary completeness

- Economic analysis of implementation costs and return on investment

- Security and privacy analysis including threat modeling and cryptographic enforcement

The federated governance approach enables manufacturing enterprises to govern IoT data at scale while preserving the operational autonomy required for production continuity. By enforcing governance at architectural boundaries rather than within operational systems, the model respects the reality that manufacturing cannot centrally control all data sources while still achieving trust, reuse, and accountability across domains.

Future work will explore empirical validation through case studies and controlled experiments, formal verification of governance properties, AI-driven governance enhancements, federated learning for quality prediction, and blockchain-based audit trail immutability. As IoT platforms continue to scale in scope and complexity, architectural frameworks that balance operational autonomy with centralized compliance will become essential to extracting value from industrial data at enterprise scale.